\documentstyle[12pt,aaspptwo,flushrt]{article}
\tighten

\begin{document}
\def\puncspace{\ifmmode\,\else{\ifcat.\C{\if.\C\else\if,\C\else\if?\C\else%
\if:\C\else\if;\C\else\if-\C\else\if)\C\else\if/\C\else\if]\C\else\if'\C%
\else\space\fi\fi\fi\fi\fi\fi\fi\fi\fi\fi}%
\else\if\empty\C\else\if\space\C\else\space\fi\fi\fi}\fi}
\def\SP{\let\\=\empty\futurelet\C\puncspace}
\def\iras{{\it IRAS}\SP}
\def\oi{{\it Optical-IRAS\SP}}
\def\eg{{\it e.g.\/\rm,\ }}
\def\etal{{\it et al.\/}\ }
\def\kms{km~s$^{-1}$\SP}


\def\fighommmag{1}         
\def\fighomwmag{2}         
\def\figtoymodel{3}        
\def\figtoybiasm{4}        
\def\figtoybiasw{5}        
\def\figbias{6}            
\def\figrealbias{7}        
\def\figbiasred{9}         
\def\figbiasomega{8}       
\def\figbiasseven{10}      
\def\figseven{11}          
\def\figbiaswill{12}       
\def\figwillcorr{13}       
\def\figvpecobssc{14}      
\def\figbiassc{15}         


\def\tabselection{2}  
\def\tabmcparell{3}   
\def\tabmcpar{1}      


\titlepage

\title{Determination of Malmquist Bias and Selection Effects
from Monte-Carlo Simulations}

\author {Wolfram Freudling}
\affil {Space Telescope--European Coordinating Facility,
European Southern Observatory, Karl Schwarzschild Str. 2, D--85748
Garching b. M\"unchen, Germany}

\author {Luiz N. Da Costa}
\affil {Institut d' Astrophysique, 98 bis Boulevard Arago, F75014,
Paris, France and Observatorio Nacional, Rua Gen. J. Cristino 77,
S\~ao Cristov\~ao, Rio de Janeiro, Brazil}

\author {Gary Wegner}
\affil {Dept. of Physics and Astronomy, 6127 Wilder Lab., Dartmouth College,
Hanover,
NH 03755 and Astronomisches Institut, Ruhr-Universit\"at Bochum, D-44780
Bochum, Germany}

\author {Riccardo Giovanelli and Martha P. Haynes}
\affil {Center for Radiophysics and Space Research
and National Astronomy and Ionosphere Center\altaffilmark{1},
Cornell University, Ithaca, NY 14953}

\altaffiltext{1}{The National
Astronomy and Ionosphere Center is operated by Cornell University
under a cooperative agreement with the National Science Foundation.}

\author {John J. Salzer}
\affil {Dept. of Astronomy, Wesleyan University, Middletown, CT 06457}

\vfill

\centerline{Accepted for publication in {\it The Astronomical Journal}}

\endtitlepage

\begin{abstract}{

Maps of the peculiar velocity field derived from distance relations are
affected by Malmquist type bias and selection effects.  Because of the
large number of interdependent effects, they are in most cases
difficult to treat analytically.  Monte Carlo simulations are used to
understand and evaluate these effects.  In these simulations the
``true'' spatial distribution and relevant properties of galaxies as
well as selection effects and observational uncertainties are
realistically modeled.  The results of the simulation can directly be
applied to correct observed peculiar velocity maps.  The simulation is
used to investigate biases in samples of measured peculiar velocities
by Lynden-Bell et al.  (1988), Willick (1990) and the new sample of
spiral galaxies by Haynes et al.  (1993).  Based on the results
obtained from the application of our method to toy models we find that
the method is a useful tool to estimate the bias induced both by
inhomogeneities and selection effects. This is a crucial  step  for the
analysis of the Haynes \etal sample which was selected with a redshift
dependent criterion.

 }
\end{abstract}


\section{Introduction}

Maps of the peculiar velocity field of galaxies are a powerful tool to
test scenarios for formation of large scale structure.  However, the
interpretation of such maps must take into account biases introduced by
the coupling of errors in the distance measurements and the spatial
distribution of galaxies and/or selection effects in the samples
considered. {Such biases in the derived distances are present even if the
underlying distance
relation is exactly known (which we will assume throughout this paper).
A discussion of how a bias-free distance relation can be derived from
data will be presented by Giovanelli \etal (1995).}
Biases originating from the spatial distribution
are usually referred to as Malmquist bias and to correct for it requires a
detailed knowledge of the underlying distribution of galaxies
from which the sample is drawn.  Biases introduced by the selection of
the sample, which we will call selection biases throughout this paper,
are caused by the correlation of the parameters used to select the
sample with parameters used to estimate distances.  In order to correct
for such biases, the distribution function describing relevant galaxy
properties is needed, as well as the detailed  criteria adopted
in selecting  the sample being considered.

A number of correction schemes for Malmquist bias have been suggested
by different workers (\eg Feast 1987,  Lynden-Bell et al. 1988;  Landy \&
Szalay
1992; Willick 1994).  A common feature of all these approaches is that
they require {\it a priori} knowledge of the distribution of galaxies in
space, which has to be either assumed (e.g. Lynden-Bell et al. 1988)
or, in the non-trivial case of an inhomogeneous distribution, derived
from the data before the correction for the bias can be made.
Different methods for estimating the required density distribution have
been considered.    {Dekel (1993) built a
treatment of the inhomogeneous Malmquist bias into their
reconstruction of the density field, assuming that
the distribution of galaxies
traces  the underlying matter distribution derived from peculiar velocities.}
Another approach  utilizes the self-consistently reconstructed
density field derived from redshift surveys (\eg Dekel 1994, Hudson 1994).

Biases resulting from selection effects can {in some situations} be either
comparable or larger than the Malmquist bias.  To correct for
selection biases, detailed knowledge of the selection process is
necessary.  However, even if this information is available it is difficult to
derive simple expressions for situations where the parameter used to
select the sample is not identical but correlates to one or several
parameters used to derive distances. {This is especially true
when a complex
set of different selection criteria is used to select the sample.
For example,
samples
where the selection depends explicitly on the redshift,
like the recently completed survey of }
peculiar velocities of Sc galaxies (Haynes \etal 1993, hereafter
referred to as the the Sc sample), are particularly difficult to
correct for this kind of bias.

The biases mentioned above are strongly related to the scatter of the
particular distance relation used. For example, the scatter in the
Tully--Fisher (TF) relation depends on the line widths of the galaxies
{in the sense that the scatter is large for galaxies of lower luminosity }
(Giovanelli \etal 1995).  Such an effect would be difficult to treat
analytically.  This fact and the unique selection criteria of the Sc
sample have motivated us to address the question of biases, both
Malmquist and selection biases, using a Monte-Carlo (MC) technique.

MC simulations of Malmquist bias and/or selection effects have
been used in the past by a number of workers (e.g.
Landy \& Szalay 1992, Teerikorpi 1993, Willick  1994), mainly for the purpose
of
testing analytic corrections.  In contrast, here the MC technique is
used to determine the bias at each point in space to
correct measured peculiar velocity fields.  The essence of our method is
to simulate galaxy catalogs using the spatial and parametric
distribution derived from observations, and select redshift-distance
samples similar to those currently available.  The strength of the
method is that all relevant effects can be incorporated to predict the
bias without the need for any simplifying assumptions.

In section 2 we review analytical
corrections for Malmquist bias which are used to test our MC results in
simple cases.  In section 3 a detailed description of the technique
is presented.  In section 4 we
compare the bias derived from our MC simulations with analytical
corrections and in section~5 we derive the bias field for
observational samples.   A brief summary is presented in section 6.

\section{Biases in the Peculiar Velocity Field}

Different analytical expressions to correct for the Malmquist and selection
type
bias can be found in the literature  (see e.g.  Willick
1994; and references therein). These expressions depend critically on
the properties of the scatter in the distance relation used.
For example, to discuss the bias on distances estimated from a TF
relation given by

\begin{equation}
I=a_{TF}\log w + b_{TF} \label{eq:tfdef}
\end{equation}

\noindent one can distinguish between the `direct' and `inverse' TF
relation. In equation~(\ref{eq:tfdef}) $I$ is the absolute I-band  magnitude
and $w$ is the corrected HI line width.  The `direct' relation
(called `forward' relation by Willick 1994) is derived from a least-squares fit
of $I$ as a function of $\log w$,
while the `inverse' relation is obtained from a least-squares fit of
$\log w$ as a function of $I$.  Independent of the distribution of the
scatter about the line defined by equation~(\ref{eq:tfdef}), a normal
distribution in $\log w$ {can be used as an} approximation for the inverse
relation, while a normal distribution in $I$ will be a better
approximation for the direct relation.  Although in this paper we focus our
attention on the TF relation, our conclusions
apply equally well to the $D_n - \sigma$ relation.

{A raw estimate of the distance $d_{raw}$ from a measured magnitude $m_I$ and
line width $w$ }

\begin{equation}
\log d_{raw}= 0.2 ( m_I - a_{TF} \log w - b_{TF} ) -3
\end{equation}

\noindent{is a biased estimate of the distance.}
The most general expression to correct
$d_{raw}$ obtained with the direct TF relation under the assumption
that the scatter is normal in magnitudes is

\begin{equation}
d_{cor}={{\int_0^\infty}r^3n(r)exp\bigl(-{[ln(r/d_{raw})]^2\over2\Delta^2}
\bigr)dr  \over
{\int_0^\infty}r^2n(r)exp\bigl(-{[ln(r/d_{raw})]^2\over2\Delta^2}\bigr)dr  }
\label{eq:malmgen}
\end{equation}

\noindent where $d_{cor}$ is the corrected estimated distance, $\Delta^2$
is the variance in the logarithm of the distance, and $n(r)$ is
proportional to the space distribution of galaxies along the line of sight
{(\eg Willick, 1994). }

Landy \& Szalay (1992) have suggested a different expression given by

\begin{equation}
d_{cor}= d_{raw} \exp\bigr({\Delta^2\over2} {h[\ln d_{raw} +
\Delta^2]\over h[\ln d_{raw}]}\bigl),\label{eq:malms}
\end{equation}

\noindent where $h(d)$ is the  {\sl observed} distribution of galaxies along
the line of sight.
As pointed out by Teerikorpi (1993), this expression is applicable when
distances
have been derived using the inverse TF relation.

Both expression (\ref{eq:malmgen}) and (\ref{eq:malms}) can be easily
applied for the homogeneous case.  They can also be easily
applied to areas of the sky where the distribution of observed galaxies can
be adequately described by one dimensional functions $h(d)$ or $n(r)$.
However,  if
larger regions of the sky are considered, variation with the direction in
the sky must be taken into account.  For example, estimates for $n(r)$ can be
obtained from redshift surveys taking into account expected peculiar velocities
(Hudson 1994).

Willick (1991) also presented a formal framework to correct for
selection biases which is similar to the corrections for Malmquist bias
given in equation~(\ref{eq:malmgen}). {Unfortunately,
his expressions cannot directly be applied to a sample selected as the
Sc sample}.

In the analysis of the Sc sample we expect significant contributions
from both Malmquist bias and selection effects due to the selection
criteria adopted.  This has prompted us to develop the MC approach
presented below.  In section~4 we compare the biases of  the analytical
expressions given in this section to results from our MC
technique discussed below.

\section{The Monte Carlo Simulation}

\subsection{Method}

The ultimate goal of the method is to create simulated redshift --
distance surveys with the same properties and thus the same biases as in
existing observed samples.  To achieve this, we have to simulate properties
like the spatial distribution, magnitude -- diameter function, color
relations, distance relations and their respective scatter properties.
In addition, the selection
criteria and completeness of the observed samples
under consideration have to be taken into account.

For the spatial distribution of galaxies, we use the density field
derived  from redshift surveys
corrected for peculiar velocities (see below).
The derived density field depends on the assumed
$\beta= \Omega/b^{0.6}$, where $\Omega$ is the density parameter and
$b$ is the linear biasing of galaxies relative to matter.  In this
paper, we assume that $\beta=1$.  It should be emphasized that the
density field and therefore the simulation is only
slightly model dependent.

We use this density field to create a large number of artificial galaxies,
to which we assign properties similar to those of the observed population
of galaxies.  We then simulate the selection of catalogs from which
observational samples are drawn.  Subsequently, the observational sample is
drawn from the simulated catalog, following the same selection criteria
used for existing surveys, including any known incompleteness of the
observational sample.  Although below we assume that this incompleteness
depends
only on the magnitudes, this can be easily generalized to include
incompleteness as a function of direction in the sky.  Finally, estimated
distances are derived from the assumed distance relation.

The final sample, which contains both the true and the estimated distances,
can now be used to compute the bias field in estimated distance or redshift
space.  For each artificial galaxy, the bias is computed by subtracting
from the estimated distance the known ``true'' distance of galaxies.  This
bias affects not only the position of the galaxies but also its estimated
peculiar velocity which can also be corrected using the same bias
correction.  Since the artificial catalogs of galaxies are large, the
uncertainty in the determination of the biases can be reduced by binning
them on a grid either in redshift or in estimated distance space.  This
average bias can then be applied to observed samples selected like the
simulation.  The raw estimated distance can be used to select the
appropriate bias from the simulation and the estimated distance can be
corrected for it.  Peculiar velocities are subsequently computed from the
bias corrected distances.

{A summary of the items and parameters that represent
 the built-in assumptions for the estimate of biases are
the parameters describing the luminosity function, the
TF relation and the color-line width relation, the density distribution
of galaxies, the peculiar velocities of galaxies, and the detailed selection
and completeness of the observed sample. }
The main advantage of this approach is the flexibility it provides to
evaluate a variety of effects and their impact on the estimated
distances, including the scatter in the various relations used, the
properties of the distance relations and the selection and
incompleteness of observational samples.

\subsection{Procedure}

\subsubsection{Creating the 3d Galaxy Distribution}

To prepare the MC simulation, either
a homogeneous  distribution of galaxies or a distribution modeled
after the local universe is used to derive a density field
in real space. For the latter, the \oi model without bias of
Freudling \etal (1994) is used. This model uses the
redshift distribution of optically selected galaxies in the galactic
caps ($b<-30^\circ$ or $b>40^\circ$), the 1.9Jy \iras sample in regions
$|b|>5^\circ$, and interpolates the density field over the galactic
plane  ($|b|<5^\circ$). The positions of the galaxies are computed
iteratively from the redshifts, using the linear approximation to
estimate the peculiar velocity of each galaxy in each iteration.
The final output is a self-consistent density field, which is
stored on a Cartesian grid of 500\kms cell
size. This final density field  is shown by Willmer \etal (1995).

The first step in the MC simulation is to create a list of
galaxy coordinates which distributes galaxies randomly but with
density fluctuations identical to those in  the \oi model.
This is accomplished by interpreting the local density  of the model
as the probability $p$ to draw a galaxy at that location. The x, y and z
components of the position of potential galaxies are randomly drawn.
The object is then rejected with a frequency proportional to $1-p$.

Finally, each galaxy is given a redshift computed from its distance, the
velocity field of the \oi model, and an assumed Gaussian random component.
The distances are assumed to be proportional to the unperturbed redshift in
the CMB restframe  and are expressed in km/sec throughout this paper.
For computational efficiency, the peculiar velocity vector is computed only
once for each cell on the density mesh, and all galaxies within each cell
are given the same peculiar velocity field.  Subsequently, a Gaussian
random component to the redshift of $\sigma_f=150$\kms is added,
representing an estimate of the cosmic velocity dispersion.

\subsubsection{Galaxy Properties}

Most observational maps of peculiar velocities (e.g.  Willick 1991,
Courteau \etal 1993, or the Sc sample) are originally drawn from
diameter-limited catalogs such as the UGC or the ESO-Uppsala catalogs,
and magnitude or other criteria are only imposed on the galaxies which
already have been pre-selected by the criteria imposed in the parent
catalog.  In order to incorporate such selection procedures into the
MC simulation, galaxies distributed according to some density
field must be given diameters, magnitudes and other
parameters relevant for the distance relation being used.  Sodr\'e \&
Lahav (1993) have investigated the joint diameter-magnitude
distribution function, and have parameterized it in the following way:

\begin{equation}
\Psi(D,B) \propto \exp\bigl[-{D\over D^{\star}}-{(b- a\log(D)-B)^2\over
2\sigma_M^2}\bigr] \label{eq:exp}
\end{equation}

where $B$ is the blue absolute magnitude computed from $B_T$, $D$
is the diameter of the galaxy, and the parameters
$a$, $b$, $D^\star$ and $\sigma_M$ have been derived by  Sodr\'e \&
Lahav.
We use such a distribution to assign both diameters and blue magnitudes to
the artificial galaxies.  For the simulation of the Sc sample, the
parameters derived by Sodr\'e \& Lahav for their ``S+I'' sample with three
different velocity models were averaged. The resulting parameters used are
$D^\star=2231$\kms arcmin, a=-4.87, b=-0.05.

The properties of each galaxy are assigned as follows.  First the
diameters in \kms arcmin are drawn from an exponential distribution function
$\propto\exp(D/D^{\star})$ (see equation~[\ref{eq:exp}]).  For a given
diameter, we derive  the most probable value for the blue absolute
magnitudes $B$ from the linear relation between diameters and
magnitudes, $B= a\log(D)+b$, where $a$ and $b$ are the ones from the
distribution functions.  The final assigned magnitude will later be
computed by adding to $B$ a Gaussian random variable with a standard
deviation of $\sigma_M$.

Since equation~(\ref{eq:exp}) is given in the blue band, it is
necessary to relate the blue $B_T$ magnitude to the $I$ magnitude. For
that we utilize a  color-line width relation of the form
 $B_T- I = a_c \log w  + b_c$.  This together with the I-band TF
relation as given in equation~(\ref{eq:tfdef}) are used to compute  $I$
and line widths from $B$.  Preliminary relations have been derived from
the Sc data  and are given in table~\tabmcpar. As above, the scatter in
these quantities will be added in the next section.

Note that our procedure differs from that of Willick (1991), who
assumes that the galaxy properties originate from a Gaussian line width
distribution rather than a luminosity function. Magnitudes are then
computed from the widths via a given TF relation, and a normally
distributed scatter is added to the magnitudes. This results in a
Gaussian luminosity function rather than the more Schechter-like
function implicitly assumed in our approach.

\subsubsection{Correlations and Scatter}

In the previous section, the parameters $D$, $B$, $I$ and $\log w$ were
assigned to each galaxy using a distribution function for $D$ and
observed relations to compute the remaining parameters.  Up to now, we
have not considered the scatter in those relations.  However, selection
biases are produced because galaxy samples are selected by observables
which are closely correlated with the quantity to be measured.  In
particular, blue magnitudes and diameters are used later to select
samples.  Therefore, their correlation with the parameters of the TF
relation, I band magnitudes and line width, must be modeled
realistically in order to obtain reliable estimates for the biases.
The scatter in the relation used to assign $B$ from $D$
equation~(\ref{eq:exp}) was assumed to be $\sigma_M=0.709$. For the TF
relation, the scatter has been estimated to be $\sigma_{TF}=0.3$.
It should be emphasized that the observed scatter in the TF relation
and therefore the biases depend on the line width (e.g.
Giovanelli \etal 1995).  {The value of 0.3 chosen here
is the value measured from the Sc sample at a line width $\log w$ of about
2.4. }
The scatter in  the color -- line width
relation, was derived from the Sc data to be $\sigma_c=0.59$.  In
table~\tabmcpar\ we list all the parameters used in the simulations,
which are used later to determine the bias for samples of spiral
galaxies.  Finally, in the discussion below we also consider  the
uncertainty in the apparent blue magnitudes listed in the catalogs
generally used to select the sample. Here we take this scatter to be
$\sigma_m=0.3$.

Finally, we are in the position of assigning values to all of the
observable quantities relevant to our simulated samples such as $B_{\rm
scatter}$, $I_{\rm scatter}$, and $w_{\rm scatter}$. These are computed
by adding normal deviates to the corresponding quantities $B$, $I$ and
$w$ computed before.  For the inverse TF relation,  four independent
Gaussian random variables $\xi_i$ with zero mean and a $\sigma=1$ are
used in the following manner.  First four random variables $\sigma_1$ to
$\sigma_4$ are defined:

\begin{equation}
             \sigma_1= \sigma_{TF}  \cdot \xi_1
\end{equation}
\begin{equation}
             \sigma_2= \sigma_{m} \cdot  \xi_2
\end{equation}
\begin{equation}
             \sigma_3= \sqrt{\sigma_{M}^2   - \sigma_{m}^2} \cdot \xi_3
                                                \label{eq:chbeg}
\end{equation}
\begin{equation}
             \sigma_4= \sqrt{\sigma_{c}^2 - \sigma_{m}^2} \cdot \xi_4
\end{equation}

These random variable are added to the the observable quantities as
computed in the previous section in the following manner. First, a
Gaussian with a scatter $\sigma_M$ has to be added to $B$ so that the
distribution of  $B$ and $\log D$ follows equation~(\ref{eq:exp}).
Using the random variables defined above, we define

\begin{equation}
             B_{\rm scatter}= B + \sigma_2 + \sigma_3
\end{equation}

in such a way that the distribution of $B_{\rm scatter}$ satisfies
equation~(\ref{eq:exp}) and the scatter in the color relation is also
reproduced if we define $I_{\rm scatter}$ to be given by

\begin{equation}
             I_{\rm scatter}=   I   +        \sigma_3 + \sigma_4.
\end{equation}

In order to reproduce the scatter in the TF relation, we define

\begin{equation}
             w_{\rm scatter}=
             w\cdot10^{(\sigma_1+\sigma_3+\sigma_4)/a_{TF}}.\label{eq:chend}
\end{equation}

It can be  verified that this leads to a Gaussian scatter term
for $B-I$ with a dispersion of $\sigma_c$ by substituting $\sigma_1$,
$\sigma_2$ and $\sigma_3$ in the relations. This is also approximately the
scatter of the color -- line width relation, because the additional terms
added to $w$ are small compared to the scatter in the color.

Similarly, for the direct TF relation implemented by scatter in magnitudes,
equations~(\ref{eq:chbeg})
to~(\ref{eq:chend}) change to:

\begin{equation}
             \sigma_3= \sqrt{\sigma_{M}^2   - \sigma_{m}^2 -
\sigma_{TF}^2}\cdot \xi_3
\end{equation}
\begin{equation}
             \sigma_4= \sqrt{\sigma_{c}^2 - \sigma_{m}^2} \cdot \xi_4
\end{equation}

\begin{equation}
             B_{\rm scatter}= B + \sigma_1 + \sigma_2 + \sigma_3
\end{equation}
\begin{equation}
             I_{\rm scatter}=   I   + \sigma_1 +        \sigma_3 + \sigma_4
\end{equation}
\begin{equation}
             w_{\rm scatter}= w\cdot10^{(\sigma_3+\sigma_4)/a_{TF}}
\end{equation}

\subsubsection{Creating Simulated Surveys}

The next step in the MC simulation is to create galaxy catalogs similar
to existing ones which we assume to be complete within the specified
criteria.  This step is necessary to simulate the exact selection
procedure of the observer.  For example, a `magnitude-limited' sample
drawn from the UGC catalog is not identical to a truly magnitude
limited sample because the inclusion of a galaxy in the UGC already
depends on a diameter selection threshold.  Since all the observables
of the artificial galaxies are known, this is an easy step.  For
example, for the UGC we select galaxies with diameters larger than 1
arcmin or brighter than 14.5 magnitudes.

Next, ``observed samples'' of galaxies are drawn from the simulated
catalogs.  It is here that we can take full advantage of the MC
approach since  besides the known selection effects (e.g. on magnitudes
and sizes of galaxies), even complicated selection criteria, such as
imposed redshifts cutoffs, variations with direction on the sky and
combination of different samples can be incorporated into our
simulations. In addition,  observational samples are characterized by
an incompleteness function that must also be taken into account in
building our simulated samples. In general, we assume that the samples
are incomplete as a function of apparent blue magnitude only. The
incompleteness of each sample was taken into account in generating our
simulated samples.

Finally,  peculiar velocities are computed from the estimated distances
obtained from the TF relation and observed magnitudes, widths and
redshifts.

\subsubsection{Computation of Bias}

The peculiar velocity of a galaxy is defined as $v_{pec}=
v_r-d_{true}$, where $v_r$ includes the recessional velocity  and the
streaming motion at its location, and $d_{true}$ is the true velocity
of the galaxy. This quantity is estimated from
measurements of the redshift $cz$ and the distance $d_{raw}$ as
$v_{pec,m}=cz-d_{raw}$, where the measured redshift includes the cosmic
random component $\sigma_f$ and the estimated distance a large measurement
uncertainty. These random components lead to a bias of the measured
peculiar velocity relative to $v_{pec}$.

The final step in our MC simulation is to estimate these biases.  This
is done by computing raw estimates of the distances, corrected for
homogeneous Malmquist bias, $d_{HMB}$ and measured redshifts $cz$ from
the observed parameters, exactly in the same manner as in the
observations.  The error in the estimated distance ${\delta}_d$ is
computed for each galaxy as

\begin{equation}
             -{\delta}_d= d_{HMB}- d_{true}
\end{equation}

\noindent
 Similarly, the error in the measured redshift relative to
the underlying streaming motion and recessional velocity is computed as

\begin{equation}
              {\delta}_z= cz- v_{r}.
\end{equation}

\noindent The signs in these definitions
were chosen so that the error in the measured peculiar velocity is
$ {\delta}_v={\delta}_d+{\delta}_z$.

Biases are produced by the fact that the expectation value of these
errors at a given estimated distance or
measured redshift are not zero for an observed galaxy sample.  In the MC
simulation, the biases in
estimated distances $b_d$ or in redshift space $b_z$ are computed by
averaging the errors for individual galaxies in the sample within
spatial cubes of 500 \kms  size, i.e.

\begin{equation}
{\rm b_d} = <{\delta}_d>_{\rm cell}
\end{equation}

\begin{equation}
{\rm b_z} = <{\delta_z}>_{\rm cell}
\end{equation}

{Depending on
whether the direct or inverse TF relation was used, the averaging in cells
$<>_{\rm cell}$  was performed differently.} In case
of the direct TF relation, galaxies were assigned to spatial cells based on
the estimated distances corrected for homogeneous Malmquist bias.  This
results in a bias field which contains the non-trivial  biases not removed by
the homogeneous Malmquist correction.  In case of the inverse TF
relation, this assignment to cells was made in redshift space.

The average and root
mean square $\sigma^2_b$ of the biases for all $N$ galaxies in each
cell were estimated, homogeneous Malmquist bias corrected
distances were computed and recorded.  The resulting field of biases was
then smoothed with a Gaussian, where the $\sigma$ in grid size was chosen
to be ($\sigma^2_b/\sqrt{N}$/150 \kms )$^{2/3}$.
{The minimum smoothing applied was a Gaussian with a $\sigma$ of~1 and
the maximum $\sigma$ used was~3.}  This
results in heavy smoothing of regions with few galaxies and large
discrepancies of the biases, and less smoothing in high density regions and
regions where the biases of the galaxies agree well.
{A similar smoothing should be applied to the real data in order to
make the derived bias field fully applicable.}

These bias fields are later applied to real observations in the following
manner.  First, the raw distance estimate of real observations is
corrected for homogeneous Malmquist bias.  Next, this corrected
distance of the galaxy and its direction on the sky is used to find the
spatial cube in which the galaxy is located and the corresponding
biases are determined.  The distance bias ${b}_d$  is then
subtracted from  $d_{HMB}$ to
find a final, bias corrected estimate of the distance. Similarly, the
redshift bias ${b}_z$ is added to  the measured redshift.
The corrected distance and redshift are subsequently used to compute
the peculiar velocity of the galaxy.

\section{Simulations}

In this section, we present a series of simulations of ``observed'' peculiar
velocities distributions and compare them to the ``true'' velocities. The raw
distances are computed from the I-band TF relation
as discussed above.
The simulations differ in the assumed spatial distribution of
galaxies.  We first present a homogeneous universe and a toy model
density distribution for the purpose of demonstrating the general
behavior of the bias and testing our procedure.  Subsequently, we use a
realistic density
distribution, modeled after the density distribution derived from
redshift surveys.  Similar models will later be used to
determine biases for existing observational samples.

\subsection{Tests of the MC Method}

We begin our analysis by applying our MC method to spatial
distributions for which the expected biases can be computed
analytically.  This allows us to test the bias field determined from
the MC method.  As our first model we consider  a magnitude-limited
sample ($m_B=14.5$) drawn from a UGC-like catalog of a homogeneous
distribution of galaxies with no streaming motion. In this experiment,
the peculiar velocities of the  galaxies are exclusively due to the cosmic
dispersion $\sigma_f$.
Figures~\fighommmag\ and~\fighomwmag\ show the biases in the observed
peculiar velocities.  In this simulation, the biases in distance ${b}_d$
are computed without the HMB correction to the estimated distances, i.e.
${b}_d= <d_{raw}- d_{true}>_{\rm cell}$.

Figure~\fighommmag\ is the case corresponding to the use of the direct
TF relation.  Panel~(a) shows the biases binned in estimated
distances.  Superimposed is the bias expected from a homogeneous
distribution computed from equation~(\ref{eq:malmgen}).  It can be seen
that the bias behaves as predicted, indicating that additional
selection biases are negligible.  In redshift space (panel~b),
Malmquist-type effects result from the cosmic random component $\sigma_f$ and
produce a bias for small redshifts. {The magnitude of this bias
is directly determined by the value
of 150\kms assumed for $\sigma_f$. } The cosmic component plays the
role of the error in distance estimate. This effect contributes to a
negative bias because on the average the redshift will be
underestimated. At larger redshift, this contribution becomes small and
selection biases dominate.

In figure \fighomwmag\ we show the bias when  the inverse TF relation
is used. As before, the solid line in panel~(a) represents the expected
bias as estimated using now equation~(\ref{eq:malms}). Note that
although relatively small, biases are present even for the inverse TF
relation binned in redshift (panel~b). At small redshifts, the cause is
the same as in the case of the direct relation discussed above. At large
redshifts, selection biases are present because of the correlation
between the line width and the blue magnitudes used to select the
sample due to the   indirect selection on line widths introduced by the
color-line width relation. This is different from the assumption of
Schechter~(1980) that there are no selection effects on line widths.

In our second model, we assume spherical symmetry and a radial density
profile which simulates a void centered on the observer which is
surrounded by a region of
high density at intermediate distances. The density falls off to the mean
density
at large distances.  The density profile is shown in
figure~\figtoymodel. From this profile, the expected streaming motion
was computed in linear approximation using $\beta=1$.

In Figure~\figtoybiasm\ and \figtoybiasw, the measured peculiar
velocities are plotted for the direct and for the inverse TF
relation, respectively. In both figures, the measured peculiar velocities and
their
biases are plotted as a function of estimated distances and redshifts.
Again, the biases do not include the HMB correction and the analytical
solutions computed from equations~(\ref{eq:malmgen})
and~(\ref{eq:malms}) are superimposed on the MC results.

Comparing  figure~\figtoybiasm\ and~\figtoybiasw\ one can clearly see
the advantage of studying the  peculiar velocities in redshift space.
{This has previously been recognized by Willick (1994). }
Figure~\figtoybiasm\ shows that although selection effects are
important, they are almost independent of density enhancements. This is an
important property, because in this case
biases do not produce any apparent infall.  More importantly
figure~\figtoybiasw\ shows that in the case of the inverse TF binned in
redshift space, the biases are smaller than in all other combinations.

The good agreement between the analytical curves and both the
homogeneous and inhomogeneous bias recovered utilizing our methodology
demonstrates that our MC technique is reliable.

\subsection{Realistic Model}

In order to deal with real  samples, it is necessary to apply our
simulations to a density field which adequately represents the true
distribution of galaxies. As discussed above, we utilize the \oi
density field for this purpose.  In contrast to the simple cases
considered earlier, the
bias $b_d$ in this section is computed from the HMB corrected
distance estimates $d_{HMB}$.

The resulting velocity biases are shown in figure~\figbias\ as vectors
along the line of sight at their $d_{HMB}$ positions for the case of a
magnitude limited sample (m=14.5) and a TF relation with scatter
distributed in magnitudes. For clarity, the scale of the bias vectors
is  five times the scale of the distances.  This plot corresponds to a
peculiar velocity field an observer would derive {\sl if there were no
true peculiar velocities} and the observer computed peculiar velocities
from estimated distances corrected for HMB.    It can be seen that the
effect of the bias is  to produce artificial infall into high-density
regions.  For instance this effect can be seen in the central panel
near the locations of the Great Attractor ($SGX\approx-5000$,
$SGY\approx+2000$), and the Pieces-Perseus complex ($SGX\approx+3000$,
$SGY\approx-3000$).

The similarity of this map with maps of the predicted peculiar
velocities from redshift surveys such as the one presented by Freudling
et al.  (1994) is evident.  Note that these strong local biases average
out if binned into homogeneous Malmquist bias corrected estimated
distance bins.  This is shown in Figure~\figrealbias.  This shows that
the HMB correction works well globally, but fails locally because of
the large density variations.

Above, we noted that the bias field in estimated distance space looks
similar to peculiar velocity maps predicted from the \oi\ model
reconstruction from redshift surveys. To quantify this impression, we
present in Figure~\figbiasomega\ the apparent velocities shown in
figure~\figbias\ and generated by the bias versus the velocities
predicted from the \oi model.  It is such a plot which in principle can
be used to derive a value for $\beta$.  The comparison of the
distribution of points with the lines for different $\beta$ suggests
that from the data one would derive a $\beta$ in the interval  0.2-0.5
even when the true peculiar velocities are zero, i.e.  the true
$\beta\approx0$.  This demonstrates that $\beta$ cannot be derived from data
which have been corrected only for HMB.

In the previous section, it was argued that distances estimated with
the inverse TF relation are not dominated by Malmquist bias if they are
plotted as a function of redshift. Instead, such distances are biased
by selection effects (see Figure~\fighomwmag). In order to demonstrate
this difference, we plot in Figure~\figbiasred\ the biases for a model
identical to the one shown in Figure~\figbias\ but with the scatter in
the TF relation taken to be in the line width, and the biases are shown
in redshift space. It can be seen that the biases show a completely
different behavior than those in estimated distance space shown in
Figure~\figbias. They are mainly a function of distance, and do not
concentrate around high density regions. For the chosen magnitude
limited sample of 14.5, the biases are significantly smaller than the one seen
in figure~\figbias.

\section{Applications}

While the purpose of the previous discussion was to understand the
properties of Malmquist bias in combination with selection effects, the
purpose of this section is to predict the magnitude of these effects
for real  surveys.  Therefore we create simulated redshift-distance
surveys which correspond to samples actually observed.  A summary of
the surveys considered and their selection criteria  is listed in
table~\tabselection.  The samples considered are those of Lynden-Bell
\etal (1988, hereafter 7S), Willick (1990, hereafter W90) and the new
Sc sample.

\subsection{7S Sample}

The parameters for the simulation of the 7S sample  of elliptical
galaxies  are given in table~\tabmcparell\ and the incompleteness as a
function of magnitude was taken from Faber \etal  (1989). Since the TF
relation is substituted by a $D_n-\sigma$ relation (see
table~\tabselection), the  color--line width relation must also be
replaced by the equivalent relation.  We could use a relation between
$D-D_n$ and $\sigma$, derived from the data,  which plays the same role
as the color-line width relation and allows for possible correlation
between the parameters used to select the sample with both of the
quantities of the distance relation. No significant correlation
between $D-D_n$ and $\sigma$ was found, therefore $D_n$ was computed
from $D$ by adding the constant and Gaussian scatter given in
table~\tabmcparell.

In figures~\figbiasseven, the derived bias field for the 7S sample is
shown.  The biases are again shown in estimated HMB corrected distances
space. The biases
produce some apparent infall into the `Great Attractor' (at
$SGX\approx-4200$\kms, $SGY\approx+800$\kms, $SGZ\approx-700$\kms), but
the effect is not much stronger than in other parts of the sky.  In
order to address the question whether such biases could produce the
velocity field as suggested by 7S, we bin in figure \figseven\ the
biases as a function of the cosine of the angular distance from the GA
in several distance shells.  {The binning and presentation corresponds
exactly to the one used by 7S} in their figure~5. It can be seen that the
biases binned in such a manner are independent of the angular distance
from the GA, unlike the behavior the 7S find in their data.  We
conclude that residual biases do not
significantly affect the conclusion of 7S regarding the existence of a large
infall towards the GA within the volume considered.

\subsection{W90 Sample}

In order to estimate the incompleteness of the W90 sample, we have
binned the data in  0.5 magnitudes intervals and compared the
resulting counts to those obtained in the CGCG (Zwicky \etal 1961-68) catalog.
This
incompleteness was imposed on the simulated sample.  Subsequently, the
biases were computed using the direct TF relation as discussed in
section~3.3.  It should be noted that the  TF relation we use differs
from that assumed by Willick~(1990).  In principle we could use the
same quadratic relation, but we believe that the major source of
uncertainty stems instead  from the lack of detailed information
regarding the properties of the scatter in the TF relation.
{In particular, the sample contains galaxies as faint as magnitude 16.5, but
the number of galaxies in each magnitude interval is small. This leaves the
possibility of substantial unrecognized systematic effects in the
incompleteness of the sample.}

In figure~\figbiaswill, the biases for the simulated W90 sample is
shown.  The figure shows the biases both as a function of redshift and
as a function of $d_{HMB}$.  The biases are shown separately for
models with  no true peculiar velocities and with a  velocity
field given by the \oi model.  The biases are essentially independent of
the assumed velocity model.  A comparison of the two panels in
figure~\figbiaswill\ shows that for a sample selected like  W90, the
biases out to 5000\kms  are smaller if the peculiar velocities are
plotted as a function of $d_{HMB}$  rather than redshifts.   Beyond
that distance, the bias is similar in amplitude but with opposite
sign.

In figure~\figwillcorr, the W90 data are corrected for the bias shown
in figure~\figbiaswill\ and binned in bins of 500~\kms.  Panel~(a)
corresponds to figure~1 shown by W90, whereas panels~(b) and~(c) show the
bias corrected data as a function of redshift and distance,
respectively.  A significant infall into the Pisces-Perseus
supercluster complex can be seen both in panels~(b) and~(c).  While the
corrections (as shown in figure~\figbiaswill) are substantial, the
basic conclusion of W90 that there is a large bulk motion of the  whole
region in the direction of the Local Group cannot be ascribed to the
biases discussed in this paper.

\subsection{Sc Sample}

Samples selected using a redshift criterion such as the Sc sample are a
completely different case.  {The Sc sample also differs from the
previously discussed sample by the fact that a large fraction of all
galaxies which satisfy the selection criteria have actually been
observed. This lowers the probability for large biases due to
unrecognized systematic effects in the selection of the sample.} In
figure~\figvpecobssc\ the `measured' peculiar velocities of a sample of
galaxies selected with the Sc sample  criteria is shown as a function
of estimated distance and redshift.  As before, the estimated distances
in this this figure have been  HMB corrected.  {At small
distances  (up to about 6000km/sec) the biases in addition to the HMB
are small.  This reflects the changing diameter limit as a function of
redshift adopted in selecting the sample. } On the other hand, at large
distance the peculiar velocities are  strongly affected by the redshift
cutoff which leads to the depopulation of galaxies with positive
peculiar velocities. The full bias map which we will actually use in
our analysis of the Sc sample is shown in figure~\figbiassc, again
plotted against HMB corrected distances.

\section{Summary}

We have presented a new approach to correct observed peculiar velocity
fields.  The Monte-Carlo   approach we have adopted  allows us to
simultaneously include all known sources of biases in a realistic way.
The resulting simulated biases  can be applied to existing observed
samples  with complex but well-defined selection criteria, such as the
new Sc sample.

We have tested the MC simulation  in simple cases, for which bias
correction schemes are  known, and recovered the expected biases with
high accuracy. We have also applied our technique to real samples and
computed the expected biases.  We found that the results presented by
Lynden-Bell \etal are virtually free of biases. On the other hand, the
data as presented by Willick (1990) are significantly biased, but
the biases do
not affect his main conclusion. Finally,  we derived the expected bias
for the new Sc sample of Haynes \etal (1993) that   will be used in
future analysis of this sample.

\bigskip

{\bf Acknowledgments} The sample of W90 was kindly provided by Jeff
Willick. GW acknowledges partial support by National Science Foundation
Grant AST93-47714 and the Alexander von Humboldt Stiftung.

\newpage



\noindent{\bf Figure Captions}

\bigskip

\def\caphommmag{\noindent{\bf Figure \fighommmag}: Biases of an homogeneous
distribution of galaxies observed with 14.5 magnitude limit drawn from the
UGC catalog.  The scatter of the TF relation of this sample is distributed
in m, corresponding to the direct TF relation.  The velocities were binned
in 500\kms bins in estimated distance $d_{\rm raw}$ in panel~(a) and
redshift $cz$ in panel~(b).  The units for distances, redshifts and biases
are km/sec.  The solid line in the panel~(a) is the homogeneous Malmquist
bias computed from equation~(\ref{eq:malmgen}).  } \caphommmag

\def\caphomwmag{\noindent{\bf Figure \fighomwmag}: Same figure as
Figure~\fighommmag, but with the scatter in the TF relation distributed in
$\log w$, corresponding to an inverse TF relation.  The solid curve in
panel~(a) is the expected bias for the observed distribution of galaxies
computed from equation~(\ref{eq:malms}).  } \caphomwmag

\def\captoymodel{\noindent{\bf Figure \figtoymodel}: Density profile and
peculiar velocities of the toy model described in the text.  } \captoymodel

\def\captoybiasm{\noindent{\bf Figure \figtoybiasm}: Observed peculiar
velocities (panels~a and~c) and biases (panels~b and~d) from the toy
model for the direct TF relation.  Panels~(a) and~(b) show the velocities
and biases binned in estimated distances $d_{raw}$, whereas panels~(c)
and~(d) show the them binned in redshift.  The dashed lines in panels~(a)
and~(b) are the expected homogeneous Malmquist bias.  The solid curve in
panel~(b) is the bias as computed from equation~(\ref{eq:malmgen}), using
the known density profile as shown in figure~\figtoymodel.  Note the strong
bias for the redshift binning (panel~d).  } \captoybiasm

\def\captoybiasw{\noindent{\bf Figure \figtoybiasw}: Observed peculiar
velocities and biases from the toy model for the inverse TF relation.  The
presentation is the same is in figure~\figtoybiasm.  The solid curve in
panel~(b) is the bias as computed from equation~(\ref{eq:malms}), using the
observed distribution of galaxies derived from the data.  Note that there
is some bias even for the redshift binning (panel~d).

} \captoybiasw

\def\capbias{\noindent{\bf Figure \figbias}: Biases of a 14.5 magnitude
limited sample shown as a function of  HMB corrected positions. The
biases were computed using the direct TF relation and are shown as
vectors at their  positions  in Supergalactic Cartesian coordinates.
This plot corresponds to a peculiar velocity field an observer would
derive {\sl if there were no true peculiar velocities} and the observer
computed peculiar velocities from estimated distances corrected for
HMB. The units for distances
are km/sec.  The scale of the velocity vectors is five times the scale of
distances. } \capbias

\def\caprealbias{\noindent{\bf Figure \figrealbias}: The biases shown
in figure~\figbias\ binned in HMB corrected estimated distances bins.
Note the very small amplitude of the bias if averaged over the whole
sky.} \caprealbias

\def\capbiasomega{\noindent{\bf Figure \figbiasomega}: Estimated HMB
corrected peculiar velocities for a model without true peculiar velocities
plotted versus the peculiar velocity as predicted by the \oi model for
$\beta=1$.  If no biases were present, the two quantities should not
correlate.  The observed correlation between estimated and predicted
velocities could be mistaken as evidence for gravitationally induced
streaming motion if biases are not taken into account.  The lines show
expected correlations for several different values of $\beta$.  The
observed correlation is similar to the one expected for values of $\beta$
between 0.2 and 0.5.} \capbiasomega

\def\capbiasred{\noindent{\bf Figure \figbiasred}: Biases of distances
derived from the TF relation for a 14.5 magnitude limited sample shown in
redshift space.  The bias map shown here is again plotted as a function of
Supergalactic coordinates.  The velocity vectors are five times the scale of
distances.  This is to be compared to figure~\figbias, which shows a
similar plot shown in corrected estimated distance space, which has a
completely different behavior of the biases.  } \capbiasred

\def\capbiasseven{\noindent{\bf Figure \figbiasseven}: Simulated biases for
the 7S sample, binned in estimated distance space.  The scale of the arrows
is 5 times the scale of distances.  }\capbiasseven

\def\capseven{\noindent{\bf Figure \figseven}: The unsmoothed biases from
figure~\figbiasseven\ plotted as a function of the cosine of the angular
distance from the Great Attractor in 3 estimated distance ranges, similar
to figure~5 in 7S.  It can be seen that conclusions by 7S based on this
plot are {\sl not} subject to biases.}\capseven

\def\capbiaswill{\noindent{\bf Figure \figbiaswill}: Biases computed for
the simulated sample of Willick (1990).  The panel~(a) shows the biases as
a function of redshift.  Panel~(b) shows the biases as a function of HMB
corrected distances.  The error bars show the biases computed for a model
with a velocity field as described in the text.  The solid points show the
biases for a model without any `true' velocities.  The solid points are
offset horizontally from their true position by 100 \kms in order to make
them easier to distinguish from the error bars.  }\capbiaswill

\def\capwillcorr{\noindent{\bf Figure \figwillcorr}: The W90 sample
corrected for biases and binned in bins of 500~\kms.  Panel~(a) shows the
raw data as a function of redshift in the CMB restframe.  Panel~(b) shows
the corrected peculiar velocities binned in redshift bins of 500\kms size.
Panel~(c) shows the corrected data binned in equivalent estimated distance
bins.  }\capwillcorr

\def\capvpecobssc{\noindent{\bf Figure \figvpecobssc}: HMB corrected
estimates of the peculiar velocities of a sample of galaxies selected like
the Sc sample, binned in  estimated distances (panel~a) and redshifts
(panel~b).  The small points indicate individual galaxies, whereas the
large points are averages in bins of 500\kms.  The dashed line is the
computed homogeneous Malmquist bias.  The most prominent feature when
plotted as a function of the estimated distance is the redshift cutoff at
7500 \kms , which has the effect that galaxies with a redshift $>7500$ \kms
are only included if they have negative peculiar velocities.
}\capvpecobssc

\def\capbiassc{\noindent{\bf Figure \figbiassc}: Bias map for the simulated
sample of Haynes et al. (1994). The unit of the velocity vectors is identical
to the
unit of distances.  The vectors are plotted at the position of the estimated
distances corrected for homogeneous Malmquist bias.  }\capbiassc

\newpage

\renewcommand\baselinestretch{1}
\baselineskip0.1cm

\begin{table*}
\caption[]{Model Parameters for Spiral galaxies}

\begin{flushleft}
\begin{tabular}{cr}
\hline
              Luminosity & function  \\
\hline
$D^*$ \kms                 & 2231    \\
a                           & -4.87   \\
b                           & -0.05  \\
$\sigma_M$                  & 0.709  \\
\hline
              TF &  relation       \\
\hline
$\sigma_{TF}  $             & 0.3        \\
$a_{TF}$                    & -7.48     \\
$b_{TF}$                    & -2.53           \\
\hline
$\sigma_{m}$                & 0.3       \\
\hline
$a_c$                       & 1.78          \\
$b_c$                       & -2.54          \\
$\sigma_c$                  & 0.59       \\
\end{tabular}
\end{flushleft}

\end{table*}

\begin{table}
\caption[]{Selection Criteria of Samples}

\begin{flushleft}
\begin{tabular}{cccccccc}
Sample & catalogs & type & area & mag. & sizes & redshifts  \\
\hline
Sc     & UGC &  Sc  & $|b|>10^\circ$ &      & $5'>d>2.5'$ & $<3000$ \\
       & ESO &      &                &      & $5'>d>1.5'$ & $ 3000$ to $5000$
\\
       & MGC &      &                &      & $5'>d>1.3'$ & $ 5000$ to $7500$
\\
\hline
 7S    & RC2 & E & $|b|>2.5^\circ$ &  $B_T^o<13.0$    & $d>2'$ & $<8000$ \\
       & RSA &      &                &      &                       \\
       & UGC &      &                &      &                       \\
       & ESO &      &                &      &                       \\
\hline
Willick   & UGC & Spirals &  $70<l<160^\circ$& $<16.5$ & $ d>0.6'$&       & \\
           &CGCG&         &   $b<0$    &      &                      & \\
\hline
\end{tabular}
\end{flushleft}

\end{table}

\begin{table*}
\caption[]{Model Parameters for E + S0}

\begin{flushleft}
\begin{tabular}{cr}
\hline
              Luminosity & function  \\
\hline
$D^*$ \kms                      & 2489  \\
a                               & -4.62 \\
b                               & -1.66 \\
$\sigma_M$                      & 0.396 \\
\hline
             $D_n-\sigma$ &  relation                 \\
\hline
$\sigma_{Dn-\sigma}  $          &   0.6     \\
$a_{Dn-\sigma}$                 &   1.2     \\
\hline
$\sigma_{m}$                    &   0.3     \\
\hline
$D-D_n$                         &   0.05         \\
$\sigma_{D-D_n}$                &   0.5     \\
\end{tabular}
\end{flushleft}

\end{table*}

\textheight=9.4in

\onecolumn
\eject
\includegraphics{malmps/bin_homm14_short.ps}
\phantom{x}\vfill\phantom{x}
\vskip1truecm\caphommmag
\newpage

\onecolumn
\eject
\includegraphics{malmps/bin_homw14_short.ps}
\phantom{x}\vfill
\caphomwmag
\newpage

\onecolumn
\eject
\includegraphics{malmps/profile.ps}
\phantom{x}\vfill
\captoymodel
\newpage

\onecolumn
\eject
\includegraphics{malmps/toym.ps}
\phantom{x}\vfill
\captoybiasm
\newpage

\onecolumn
\eject
\includegraphics{malmps/toyw.ps}
\phantom{x}\vfill
\captoybiasw
\newpage

\onecolumn
\eject
\includegraphics{malmps/real_m_14_bias.ps}
\phantom{x}\vfill
\capbias
\newpage

\onecolumn
\eject
\includegraphics{malmps/realbias.ps}
\phantom{x}\vfill
\caprealbias
\newpage

\onecolumn
\eject
\includegraphics{malmps/omega.ps}
\phantom{x}\vfill
\capbiasomega
\newpage

\onecolumn
\eject
\includegraphics{malmps/real_w_14_bias.ps}
\phantom{x}\vfill
\capbiasred
\newpage

\onecolumn
\eject
\includegraphics{malmps/ell_m_bias.ps}
\phantom{x}\vfill
\capbiasseven
\newpage

\onecolumn
\eject
\includegraphics{malmps/seven.ps}
\phantom{x}\vfill
\capseven
\newpage

\onecolumn
\eject
\includegraphics{malmps/biaswill.ps}
\phantom{x}\vfill
\capbiaswill
\newpage

\onecolumn
\eject
\includegraphics{malmps/data_will_cmb.ps}
\phantom{x}\vfill
\capwillcorr
\newpage

\onecolumn
\eject
\includegraphics{malmps/vpecobssc_mb.ps}
\phantom{x}\vfill
\capvpecobssc
\newpage

\onecolumn
\eject
\includegraphics{malmps/real_m_sc_bias.ps}
\phantom{x}\vfill
\capbiassc
\newpage


\bigskip








\begin{references}

{

\reference  Bothun, G.D.,  \& Cornell, M.E., 1990, AJ 99, 1004.


\reference  Courteau, S., Faber, S.M., Dressler, A., Willick, J.A.,
            1993, ApJ 412, 51. 

\reference  Dekel, A., 1993, in Observational Cosmology, ASP Conference Series
Vol 51, eds G. Chincarini, A. Iovino,
            T. Maccacaro and D. Maccagni, p. 194. 

\reference Dekel, A. 1994, ARAA 32, 371.

\reference  Faber, S.M., Wegner, G., Burstein, D., Davies, R.L.,
            Dressler, A., 1989, ApJS 69, 763.

\reference Feast, M.W. 1987, Observatory, 107, 185.

\reference  Freudling, W., da Costa, L.N., Pellegrini, P.S. 1994,
            MNRS, 268, 943.

%

\reference Haynes, M.P., Giovanelli, R., Salzer, J.J., Wegner, G.,
           da Costa, L.N., Freudling, W.,  Chamaraux, P., 1993,
           BAAS, 25, 1403         

\reference Giovanelli, R. et al., 1995, in preparation

\reference Hudson,M.J.,  1994, MNRS 266, 468. 

\reference Landy,S.D.,  \& Szalay,A.S.,  1992, ApJL 394, 25L

\reference Lynden-Bell, D., Faber, S.M., Burstein, D., Davies, R.L.,
           Dressler, A., Terlevich, R.J., and Wegner, G. 1988, Ap.J., 326, 19.

\reference Nilson,P. 1973,
         { \sl Uppsala General Catalog of Galaxies}
         {Uppsala Astron. Obs. Ann.}{6}{(UGC)}

\reference Schechter,P.L. 1980, ApJ {85}, {801}

\reference Sodr\'e, Jr.L. \& Lahav, O. 1993 MNRS 260, 285.

\reference Teerikorpi, 1993, AA 280, 443.

\reference Willick, J.A. 1991, PhD thesis, {\sl University of California at
Berkeley} 

\reference Willick, J.A. 1990, ApJ 351, L5.

\reference Willick, J.A. 1994, ApJS 92, 1.

\reference Willmer C.N.A., da Costa, L.N.,  Pellegrini, P.S.,  Fairall,
           A.P.,  Latham, D.W. \& Freudling, W., 1995, AJ 109, 61.


\reference{Zwicky,F., Herzog,E., Will,P., Karpowicz,M., Kowal,C.
	 1961-1968}{\sl Catalog of Galaxies and Clusters of Galaxies
	 {\rm in six volumes}}
         {California Institute of Technology, Pasadena, (CGCG)}

}

\end{references}
\end{document}